\documentclass[10pt,twocolumn,letterpaper]{article}

\usepackage[top=0.7in,bottom=0.7in,left=0.6in,right=0.7in,columnsep=0.35in]{geometry}

\usepackage{microtype}
\usepackage{graphicx}
\usepackage{subfigure}
\usepackage{booktabs}

\usepackage{hyperref}
\usepackage{hhline}
\usepackage{amsmath,amsthm,amssymb}

\usepackage[utf8]{inputenc}

\title{Bayesian Hierarchical Bernoulli-Weibull Mixture Model for Extremely Rare Events}

\author{Yuki Ohnishi$^1$ and Shinsuke Sugaya$^2$}
\date{%
    $^1$Purdue University and%
    $^2$Bizreach, inc.\\[2ex]%
    \today
}
\begin{document}

\maketitle
\begin{abstract}
 Estimating the duration of user behavior is a central concern for most internet companies. Survival analysis is a promising method for analyzing the expected duration of events and usually assumes the same survival function for all subjects and the event will occur in the long run. However, such assumptions are inappropriate when the users behave differently or some events never occur for some users, i.e., the conversion period on web services of the light users with no intention of behaving actively on the service. Especially, if the proportion of inactive users is high, this assumption can lead to undesirable results. To address these challenges, this paper proposes a mixture model that separately addresses active and inactive individuals with a latent variable. First, we define this specific problem setting and show the limitations of conventional survival analysis in addressing this problem. We demonstrate how naturally our Bernoulli-Weibull model can accommodate the challenge. The proposed model was extended further to a Bayesian hierarchical model to incorporate each subject's parameter, offering substantial improvements over conventional, non-hierarchical models in terms of WAIC and WBIC. Second, an experiment and extensive analysis were conducted using real-world data from the Japanese job search website, CareerTrek, offered by BizReach, Inc. In the analysis, some research questions are raised, such as the difference in activation rate and conversion rate between user categories, and how instantaneously the rate of event occurrence changes as time passes. Quantitative answers and interpretations are assigned to them. Furthermore, the model is inferred in a Bayesian manner, which enables us to represent the uncertainty with a credible interval of the parameters and predictive quantities.  \end{abstract}

\maketitle

\section{Introduction}
This paper addresses the challenge of estimating the duration of time within which extremely rare events occur. In the context of web-service operation, estimating the duration of a user's lifetime on the service is often necessary to assess the service quality. For instance, the duration until a user's first action, such as the first purchase at EC or first login after registration, is a central concern of internet companies. In general, survival analysis is used widely when attempting to estimate the duration within which a certain event happens. The cases described above can be accommodated within the survival analysis framework.

However, a strong assumption about survival analysis is that it usually posits the same survival function for all subjects in which the event would occur in the long run, depending on the hazard function, which is defined as the event rate at time t conditional on survival until time t or later (i.e., $T>t$). These assumptions might affect analysis results in the following cases. One can imagine a case in which numerous users have just signed up for a certain web service, but they do not intend to use the service a lot. In such cases, a large proportion of the inactive users would never even log in the service. Therefore, it is unreasonable to assume that these inactive users are following the same survival function as the one of active users.  

To address this challenge, this study introduces a Bayesian Hierarchical Bernoulli-Weibull Mixture Model that can be applied appropriately to a setting in which a large number of events will not happen. This model presents a clear interpretation of user behavior by separating users into active and inactive users. Then it naturally models the lifetime of active users in a quantitative manner. 

Our model has been inspired by what is called the cure rate model in medical science. There is a rich literature on this model in medical trial statistics. In that sense, our methodology has little novelty. So, our contributions are mostly on the applied side. We demonstrate how naturally the cure rate model can accommodate our problem settings and present the interesting insights extracted by the model. The contributions of this paper are as follows.

\begin{itemize}
  \item A new application area to the rich literature of survival analysis, and a Bayesian Hierarchical Bernoulli-Weibull Mixture Model to evaluate the duration within which low-frequency events occur on web services.
  \item Experiments conducted on web service with real-world data, with comparisons of several models including the proposed model.
  \item Interpretations of experimentally obtained results with consideration of model uncertainty by Bayesian inference and redefines user active rate which often matters on internet services.
\end{itemize}

It is well known that capturing rare events may provide more impact than simply capturing trends: e.g., will someone get hired, will some content go viral and will a failure occur. In this sense, the hierarchical model may be used in different settings.

The remainder of this paper is structured as follows. Section 2 explains several related works of this study. Section 3 introduces some shortcomings related to classical survival analysis in this problem setting and proposes the mathematical formulation of a Bayesian Hierarchical Bernoulli-Weibull Mixture Model. How the model solves difficulties that are intractable with classical survival analysis is also explained. Section 4 describes experiments conducted with real-world data. Models are compared using WAIC and WBIC metrics. Section 5 presents discussions of the results, with responses to several research questions. Section 6 presents a conclusion to this paper.

\section{RELATED WORK}
This study examines the application for survival analysis, which is used widely in diverse scientific fields such as social sciences, marketing sciences and medical sciences. In the context of marketing science, Helsen and Schmittlein (1993) validated hazard rate models for household interpurchase times and proved the model effectiveness \cite{Helsen}. As one example of social science efforts, survival analysis was applied to detect community changes in a social network over time \cite{Tajeuna}. In the fields of medical and health sciences, some advanced survival models have attracted attentions in recent years \cite{Chen}, \cite{Yin}, \cite{Kim}. For our study, the most closely related model is a cure rate model, which incorporates a cure fraction of subjects' cancer into existing survival model. This model is becoming increasingly popular in analyzing data from cancer clinical trials. Gu (2011) proposed a new proportional odds survival model with a cured fraction using a hierarchical structure of latent factors activating cures \cite{Gu}. Our study draws an analogy between our Bernoulli-Weibull Mixture model and the cure rate model, and presents the first analysis of the duration within which low-frequency events occur on a web service based on that model.

This paper is also related to work conducted on online user behavior modeling, which has attracted growing attention in recent years. Most such works were conducted in the context of Information Retrieval (IR) \cite{Joachims} \cite{Kim14}. The approach applied in our work is similar to that described by Liu, et al. (2010), who assumed a Weibull distribution with Web page dwell time data and who analyzed implicit feedback involving the dwell time \cite{Liu}. Particularly, their analysis is conducted with non-observed data, based on the assumption that the users dwell time follows a Weibull distribution, whereas our work shows that assuming Bernoulli-Weibull mixture distribution is preferred for modeling the occurrence of very rare events with consideration of observed data \cite{Liu}.

In the area of advertisement marketing, survival analysis is also used to improve post-click user engagement on native ads. An earlier study \cite{Barbieri} was conducted via survival analysis to estimate engagement by predicting the dwelling time on a corresponding ad landing page. The author estimates the distribution of the length of time users spent on the ad and integrates it into ad ranking functions. This work is also related to ours in the sense that both try to model the duration within which internet users become involved. Generally speaking, most internet users are not very active. Therefore, a chance exists of separating the observed data into data that follow a survival function and data for which no event is going to happen in the future. This is a claim of this study. 

Next, we describe the methods and the experiments we conducted to evaluate their performance. We begin with the problem setting and formulation.

\section{PROBLEM SETTING AND FORMULATION}
This section provides details of survival analysis \cite{Ibrahim} and how it can model the expected duration within which events happen. As described herein, survival analysis works well for modeling duration of time in the context of user conversion on the internet service, and the limitations of existing methods. Then, a Bernoulli-Weibull Mixture Model is introduced to overcome challenges that the existing model cannot handle. The mixture model is extended further to a hierarchical model to evaluate the uniqueness of each user's event.

\subsection{Survival Analysis and the Motivation for a Mixture Model}
Survival analysis is used widely to analyze data in which the time until the event is of interest. The response is often designated as a failure time, survival time, or an event time. For simplicity, it is designated as an event time herein. It is noteworthy that this technique has much wider applicability, although it is always modeled as the duration of user's conversion in the present study. 

The survival analysis includes some observations of the data that are observed at time $t_i$. Standard regression procedures can be applied if no censoring is done. However, these procedures might not be inadequate for the following reasons.

\begin{itemize}
  \item Time $t_i$ is non-negative, with a skewed distribution.
  \item The probability of surviving past a certain point in time might be of more interest than the expected time of the event.
  \item The hazard function, used for regression in survival analysis, can lend greater insight into the mechanism of the event occurrence. 
\end{itemize}

We will introduce the key idea of survival analysis here. First, let $T_i$ be a continuous non-negative random variable representing an event time of user $u_i$. The distribution of $T_i$  can be characterized by the probability density function $f(t_i)$ (PDF) and cumulative distribution function $F(t_i)$ (CDF). 

\begin{equation} \label{eq:1}
  F(t_i) = Pr(T_i \leq t_i)=\int_0^{t_i} f(t)dt
\end{equation}

In survival analysis, we often specifically examine the survival function $S(t_i)$, hazard function $h(t_i)$, and cumulative hazard function $H(t_i)$. The survival function represents the probability that a subject will survive past time $t_i$. As  \ref{eq:2} shows, the survival function is the complementary cumulative distribution function (CDF). The hazard function represents the instantaneous rate at which events occur, given that no event occurred prior. The cumulative hazard function describes the accumulated risk up to time $t_i$. 

\begin{equation} \label{eq:2}
  S(t_i) = 1 - F(t_i)
\end{equation}
\begin{equation} \label{eq:3}
  h(t_i )=\lim_{t_i \to \infty}\frac{Pr(t_i\leq T_i \leq t_i+\Delta t_i | T_i>t_i)}{\Delta t_i}=\frac{f(t_i)}{S(t_i)}
\end{equation}
\begin{equation} \label{eq:4}
  H(t_i)= \int_0^{t_i} h(u)du =\int_0^{t_i} \frac{f(u)}{S(u)}du = -log(S(t_i))
\end{equation}

Several non-parametric models are used in survival analysis, but we specifically examine parametric models herein, which are more interpretable. In general, when estimating the model parameters, one must calculate the likelihood of the model. As stated before, in survival analysis, the observations of some data are observed. Let $\delta_i$ be the indicator variable that shows censoring (observed, 1; not observed: 0). The contribution to the likelihood is expressed as  \ref{eq:5} if the data are non-observed data. If otherwise, then the contribution to the likelihood is expressed as  \ref{eq:6} because event time $T_i$ must be at least $t_i$, the probability of which equals $Pr(T_i \leq t_i) = S(t_i)$. The whole likelihood of the model is calculated as \ref{eq:7}.

\begin{equation} \label{eq:5}
  \prod_{\delta_i=0} f(t_i)
\end{equation}
\begin{equation} \label{eq:6}
  \prod_{\delta_i=1} S(t_i)
\end{equation}
\begin{equation} \label{eq:7}
  L_s(\lambda, \theta) = \prod_{i=1}^{N} (f(t_i)))^{\delta_i}(S(t_i)))^{1-\delta_i}
\end{equation}

It is necessary to mention that we can assume several distributions for parametric survival analysis, such as the Exponential distribution, Weibull distribution, and log-logistic distribution. Additionally, several types of censoring exist: right-observed, left-observed, and interval observed. Although we specifically examine a Weibull distribution as survival distribution and right-observed censoring in this paper, the key ideas presented are applicable to any settings. Additional information related to survival analysis is available in the literature [13].

In the context of web service, the duration until user $u_i$ does some action we intend, might be regarded as an event time. Such an action, which might be a first purchase of an item or registration of payment members, can be called conversion. We are invariably motivated to estimate the event time until user conversion. With that information, a marketing director or a product manager of the service can draw up a marketing budget for the next period or evaluate the service quality. In such cases, survival analysis is the first option. 

An important difficulty is that, depending on the service traits, many inactive users might take almost no action on the service because they signed up merely to receive other benefits unrelated to the service. Equation \ref{eq:6} shows that existing survival analyses assume that all inactive users follow the same distribution. However, this assumption can be distorted by inactive users because the events of those users would never happen. This assumption limits the utility of classical survival analysis. To address this challenge and evaluate event duration correctly in such situations, application of a Bernoulli-Weibull Mixture Model is proposed for survival analysis of the occurrence of extremely rare events.

\subsection{Bernoulli-Weibull Mixture Model}
This subsection presents a proposal Bernoulli-Weibull Mixture Model for survival analysis. The concept is simple. Users are separated into two groups. The first group is prospective users for whom event occur in the long run. The second group is inactive users who will never take the intended action. This separation is expressed with indicator variable $Z_i$ as shown below.

\begin{equation} \label{eq:8}
Z_i \sim Bern(q)
\end{equation}
Actually, $Z_i$ is an unobserved indicator variable that is 1 if the user $u_i$ is a prospective user and 0 if the user is an inactive user. As shown in  \ref{eq:8}, it is assumed that $Z_i$ follows Bernoulli distribution with parameter $q$, which can be regarded as the probability of the user becoming a prospective user. In that sense, $Z_i$ can be regarded as a latent variable that represents whether the event will occur in the long run with probability $q$. Next, the event time can be modeled with the Weibull distribution as presented below.

\begin{equation} \label{eq:9}
  f(t_i)=\frac{\lambda t_i^{\lambda-1}}{\theta}exp\Bigl[-(\frac{t_i}{\theta})^{\lambda}\Bigr]
\end{equation}

\begin{displaymath}
  S(t_i)=
    \int_{t_i}^{\infty} \frac{\lambda u^{\lambda-1}}{\theta}exp\Bigl[-(\frac{u}{\theta})^{\lambda}\Bigr]du\quad( Z_i=1)
\end{displaymath}
\begin{equation} \label{eq:10}
  S(t_i)=1 \quad (Z_i=0)
\end{equation}

In the functions,  $\lambda$ denotes a shape parameter; $\theta$ is a scale parameter $(\lambda>0,\theta>0$). Approximately, $S(t_i)$ equals 1 when $Z_i=0 $ because we assume that the event of inactive users will never happen, which we can be expressed with CDF as follows.
\begin{equation} \label{eq:11}
    F(t|Z=0)=1-S(t|Z=0)=\int_0^t f(f|Z=0)dt \approx 0
\end{equation}

Based on this approximation, the likelihood is calculated as shown below.
\begin{displaymath}
L_m(q, \lambda, \theta)=
\end{displaymath}
\begin{displaymath}
\begin{split}
  \prod_{i=1}^{N}\Bigl[(Bern(Z_i=1|q)f(t_i))^{\delta_i}(Bern(Z_i=0|q)S(t_i |Z_i=0) \\ +Bern(Z_i=1|q)S(t_i |Z_i=1))^{1-\delta_i}\Bigr]
  \end{split}
\end{displaymath}

\begin{equation} \label{eq:12}
\begin{split}
  \approx \prod_{i=1}^{N}\Bigl[(Bern(Z_i=1|q)f(t_i))^{\delta_i}(Bern(Z_i=0|q)\\ +Bern(Z_i=1|q)S(t_i |Z_i=1))^{1-\delta_i}\Bigr]
  \end{split}
\end{equation}

\subsection{Extension to a Hierarchical Model}
We further extend the mixture model to a hierarchical model. This extension enables us to assume subjects of various types for the analysis. In various application settings, it is often the case that different subjects have different preferences and intentions. For instance, in web services, we often have access to users' demographic data such as age and gender. In such a case, it might not be reasonable to assume that behaviors of all users follow the same distribution with the same parameters, although it still stands to reason that they should be mutually related. A hierarchical model can achieve a situation in a natural way using a prior distribution in which all subjects' parameters are viewed as a sample from common distribution. This subsection explains extension of the Bernoulli-Weibull Mixture Model to the hierarchical model.

The model discussed above has parameters of three types, the probability of Bernoulli trials q, and parameters of Weibull distributions $\lambda$ and $\theta$. Assuming that users $u_i$ are categorized into categories $k_i (k_i \in K)$. We assume the same parameters for each category and assume that they are generated from common prior distributions. We set a beta prior distribution for a Bernoulli trial. A beta distribution takes two hyperparamters, $\alpha$ and $\beta$. For computational efficiency, these hyperparameters were reparameterized with $\alpha=\alpha/(\alpha+\beta)$  ,$\kappa=\alpha+\beta$. We respectively assigned a Uniform prior distribution and Pareto prior distribution to $\mu$ and $\kappa$. This reparameterization allows MCMC sampling to easily and rapidly get converged. An earlier report of the relevant literature [14] explains reparameterization and hyperprior distributions. A Bernoulli trial, which determines whether users will become active or not in this case, can be modeled as shown in  \ref{eq:13}.
\begin{displaymath}
  \alpha=\mu \kappa
\end{displaymath}
\begin{displaymath}
  \beta=\kappa(1-\mu)
\end{displaymath}
\begin{displaymath}
  \kappa \sim Pareto(0.1, 1.5)
\end{displaymath}
\begin{displaymath}
  \mu \sim Uniform(0, 1)
\end{displaymath}
\begin{displaymath}
  q_{k_i} \sim Beta(\alpha, \beta)
\end{displaymath}
\begin{equation} \label{eq:13}
    Z_i \sim Bern(q_{k_i})
\end{equation}

Parameters $\lambda$ and $\theta$ for each category should be sampled from the common prior. Herein, we assume that parameters $\lambda$ and $\theta$ are drawn respectively from a normal distribution with hyperparameters ($\mu_{\lambda}$,$\sigma_{\lambda}$) and ($\mu_{\theta}$,$\sigma_{\theta}$).

\begin{equation} \label{eq:14}
    \lambda_{k_i} \sim Normal(\mu_{\lambda}, \sigma_{\lambda})
\end{equation}
\begin{equation} \label{eq:15}
    \theta{k_i} \sim Normal(\mu_{\theta}, \sigma_{\theta})
\end{equation}

As shown in  \ref{eq:12}, the likelihood of event occurrence should be calculated considering censoring and Bernoulli trial. The likelihood of category $k$ is calculated as $L_m$ ($q_k,\lambda_k,\theta_k$ ) with the parameters for the category. The whole log-likelihood $L_h$ is calculated as the summation of log-likelihoods of the respective categories as follows.

\begin{equation} \label{eq:16}
    \L_h = \prod_{k=1}^K L_m (q_k,\lambda_k,\theta_k)
\end{equation}

\section{EXPERIMENTS}
This section presents verification of the proposed hierarchical model capacity for real-world data generated at a certain web service. An introduction of the problem setting in CareerTrek, a Japanese job search platform operated by BizReach, inc. is presented. Then available data are described. Finally, as discussed in the previous section, the performance metrics are defined and models are applied to the data. We report experimentally obtained results with comparison to each model. 

\subsection{CareerTrek}
On CareerTrek, a popular matching platform for junior job seekers and headhunters in Japan, job seekers register their own resumes and seek jobs in which they are interested. They can search independently or merely take some recommended jobs that are personalized for them each day. Headhunters can also seek prominent job candidates by searching or by personalized recommendations. The number of newly registered job seekers is growing every day along with the growth of career-change market in Japan. Analyst of this service report that the greatest interest when users sign up is in how long it takes for job seekers to receive a job offer. The event time must be estimated accurately using the latest registered data that are available, which engenders compilation of a marketing budget for subsequent periods or evaluating service quality for users.

\subsection{Experiment Setup}
This subsection presents a detailed description of our experimental setup, including data collection and evaluation methodologies, to compare the three models discussed in the previous section.

We estimate all parameters using Bayesian inference. The previous subsection introduced the likelihood of each model. A uniform distribution was used as a prior distribution unless otherwise specified. We inferred all parameters by Markov Chain Monte Carlo (MCMC) with Stan \cite{Carpenter}. Stan, a state-of-the-art platform that is useful for statistical modeling and high-performance statistical computation, offers a probabilistic programming language in which users specify log density functions and conduct full Bayesian statistical inference with MCMC sampling. The default internal algorithm is No-U-Turn Sampler (NUTS) \cite{Hoffman}, which is an extension of Hamiltonian Monte Carlo (HMC) that eliminates the need to set the problematic number-of-steps parameter. Bayesian inference is used widely to estimate the model parameters. One can estimate a Bayesian credible interval on the parameters, by sampling from the posterior distribution over the parameters, which enables us to assess model uncertainty \cite{Gelman14}.

\subsubsection{Data Collection}

\begin{table*}[t]
\caption{Summary of Collected Data}
  \label{tab:1}
\begin{tabular}{lcccc}
\hline
                    & \textbf{\#No-observed Data} & \multicolumn{1}{l}{\textbf{\#Observed Data}} & \multicolumn{1}{l}{\textbf{Average Duration}} & \multicolumn{1}{l}{\textbf{Std. of Duration}} \\ \hline
\textbf{Category 1} & 102                         & 14295                                        & 81.05                                         & 50.62                                         \\
\textbf{Category 2} & 21                          & 1215                                         & 75.92                                         & 48.47                                         \\
\textbf{Category 3} & 153                         & 47203                                        & 74.44                                         & 47.93                                         \\
\textbf{Category 4} & 1                           & 636                                          & 86.47                                         & 53.23                                         \\
\textbf{Category 5} & 13                          & 4902                                         & 78.69                                         & 53.01                                        
\end{tabular}
\end{table*}

We collected 180-day event data of a certain period of time from $T_s$ to $T_e$. There were about 70,000 users' lifetime data. All users signed up for the service between $T_s$ and $T_e$ . They were selected randomly in terms of duration. The flag of censoring and a certain category the users belong to are assigned to each user.  Some data are right-observed at time $T_e$, but no data are left-observed because we can always recognize when the user signed up, i.e., we always know the duration of observed data from  $T_s$, which means it is never left-observed. A summary of collected data is presented in Table \ref{tab:1}.

\subsubsection{Simulation Conditions}
As stated already in the previous section, the whole parameter inference was conducted by MCMC with Stan. Several MCMC parameters must be fixed before running a simulation. These parameters are presented in Table \ref{tab:2}. We used ver. 2.16.0 of Stan with the python interface: PyStan \cite{Stan}.

\subsubsection{Evaluation Methodologies}
The proposed model has a hierarchical structure and hidden variable. It is no longer a regular but a singular statistical model. In a singular statistical model, the log likelihood function cannot be approximated in any quadratic form. It is therefore inappropriate to use conventional information criteria such as AIC, BIC, TIC, and DIC for model evaluation. Leave-one-out cross-validation (LOOCV) is also a method for estimating pointwise out-of-sample prediction accuracy from a fitted Bayesian model. However, in terms of computation, it is too expensive for use in our case. The Widely Applicable Information Criterion (WAIC) and Widely Applicable Bayesian Information Criterion (WBIC) offer various advantages over classical estimates of predictive error, such as AIC and DIC. Moreover, they are less expensive in terms of computational cost than LOOCV. In our experiments, these two criteria are used for model evaluation. In the next subsection, three models, baseline survival model, Bernoulli-Weibull Mixture model and Hierarchical Bernoulli-Weibull Mixture model, are compared based on WAIC and WBIC. Detailed discussions of both criteria are presented in earlier reports of the literature  \cite{Vehtari} \cite{Watanabe10} \cite{Watanabe13_1} \cite{Watanabe13_2}.

\begin{table}[]
\caption{Simulation Parameters}
  \label{tab:2}
\begin{tabular}{cc}
\hline
\textbf{Number of chains}  & 4    \\
\textbf{Burn-in}           & 1000 \\
\textbf{Number of samples} & 4000 \\
\textbf{MCMC algorithm}    & NUTS
\end{tabular}
\end{table}

\subsection{Model Comparisons and Simulation Results}
Table \ref{tab:3} presents WAICs and WBICs with respect to three competing models. Both statistics clearly indicate that the proposed model outperformed the other baseline models, with the smallest WAIC = 6080 and WBIC = 3082. Table \ref{tab:4} presents a summary of sampling results obtained using the proposed model. Shape parameters $\lambda_k$ and scale parameters $\theta_k$ are the parameters of the Weibull distribution. The parameter of Bernoulli trial is expressed as $q_k$. The proposed model has a hierarchical structure with parameters for each category. We present the posterior Maximum a Posteriori (MAP) estimate, standard deviation, 95\% credible interval, and $\hat{R}$ for each parameter. $\hat{R}$ is the potential scale reduction factor on split chains (at convergence, $\hat{R}=1$) \cite{Gelman13}. In general, $\hat{R} < 1.1$ is a good indicator of convergence of the parameter estimation.  In that sense, all parameters for the model are converged. Further discussions are presented hereinafter.

\begin{table}[]
\caption{Model Comparison by WAIC and WBIC}
  \label{tab:3}
\begin{tabular}{lcc}
\hline
                    & \textbf{WAIC} &  \multicolumn{1}{l}{\textbf{WBIC}} \\ \hline
\textbf{Baseline model} & 6141        & 3092  \\
\textbf{Non-hierarchical model} & 6110       & 3084                    \\
\textbf{Proposed model} & 6080                         & 3082                                                              
\end{tabular}
\end{table}

\begin{table}[]
\caption{Posterior MAP, standard deviation, 95\% credible interval for the parameters of the proposed model}
  \label{tab:4}
\begin{tabular}{ccccc}
\hline
\textbf{}   & \textbf{MAP} & \textbf{post. sd} & \textbf{Credible Interval} & \textbf{$\hat{R}$} \\ \hline
\textbf{$q_1$} & 0.0152       & 0.0110            & (0.0118, 0.0251)           & 1.0095        \\
\textbf{$q_2$} & 0.0411       & 0.2024            & (0.0240, 0.8864)           & 1.0187        \\
\textbf{$q_3$} & 0.0079       & 0.0880            & (0.0066, 0.1186)           & 1.0228        \\
\textbf{$q_4$} & 0.0015       & 0.1169            & (0.0001, 0.3429)           & 1.0146        \\
\textbf{$q_5$} & 0.0040       & 0.0933            & (0.0022, 0.2555)           & 1.0168        \\
\textbf{$\lambda_1$} & 2.0167       & 0.1905            & (1.6389, 2.3865)           & 1.0030        \\
\textbf{$\lambda_2$} & 1.8029       & 0.2916            & (1.2472, 2.4047)           & 1.0241        \\
\textbf{$\lambda_3$} & 1.9200       & 0.1772            & (1.4865, 2.2411)           & 1.0077        \\
\textbf{$\lambda_4$} & 1.7754       & 0.5443            & (0.4151, 2.4842)           & 1.0086        \\
\textbf{$\lambda_5$} & 1.7684       & 0.3361            & (0.8671, 2.1048)           & 1.0142        \\
\textbf{$\theta_1$} & 97.6492      & 50.1192           & (84.7204, 162.9395)        & 1.0093        \\
\textbf{$\theta_2$} & 106.4787     & 358.2478          & (86.0139, 1357.1233)       & 1.0179        \\
\textbf{$\theta_3$} & 101.1486     & 423.8843          & (91.5024, 983.0129)        & 1.0240        \\
\textbf{$\theta_4$} & 100.8410     & 893.7816          & (17.4198, 2942.1575)       & 1.0201        \\
\textbf{$\theta_5$} & 98.2040      & 858.3811          & (59.9455, 3127.5013)       & 1.0223       
\end{tabular}
\end{table}

\section{RESEARCH QUESTIONS AND DISCUSSION}
This section presents further discussions of the sampled results for each parameter and its interpretation compared with our domain knowledge. To make the discussion more straightforward, we specifically examine the proposed Bayesian Hierarchical Bernoulli-Weibull Mixture model. Discussions of three types are presented here. First, we discuss the differences of active rates between categories and how they are estimated in a Bayesian manner. Second, a discussion of uncertainty is given. Finally, we show how to interpret the survival function and hazard function.

\subsection{Estimated Active Rate}
\begin{figure}[h]
  \centering
  \includegraphics[width=\linewidth]{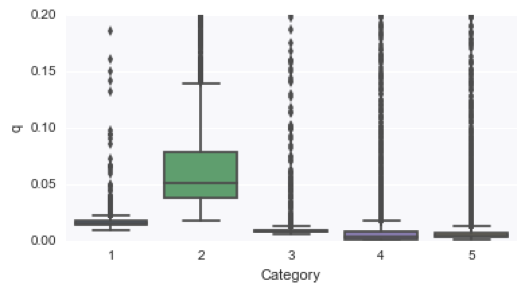}
  \caption{Posterior distribution of the active rate of each category (capped at 0.20).}
  \label{fig:fig1}
\end{figure}

The estimated active rate is one of the key results in this study. Figure \ref{fig:fig1} presents the posterior distribution of parameters q for each category. Because this parameter of a Bernoulli trial defines the latent variable $Z_i$, which indicates whether the event of the user is going to happen in the long run, this can be regarded as the Estimated Active Rate of users. This informative estimate could never be obtained without the Bernoulli assumption. As shown in Table \ref{tab:4}, category 1 has relatively sharp distribution with 95\% credible interval between 0.0118 and 0.0251. On the other hand, category 2 has a distribution with a peak at 0.0411, meaning that the event of category 2 is more likely to occur in the future. However, the distribution is relatively flat, which means that this model cannot uncertainly estimate $q_2$.

\subsection{Uncertainty of Estimated Values}
As might be apparent from Figure \ref{fig:fig1}, the distribution of $q_2$ has a flat distribution, whereas the others have sharp distributions, which indicates that the model is not very confident of its estimation because of the small size of the data in category 2. It might be too risky to choose MAP or EAP upon prediction. If one wants to estimate other generated values with these uncertain estimates (in this case, $q_2$), one might want to estimate it in a fully Bayesian manner, Monte Carlo inference, with all the information of the posterior. Full Bayesian inference involves propagation of the uncertainty in the values of parameters modeled by the posterior.

\begin{figure}[h]
  \centering
  \includegraphics[width=\linewidth]{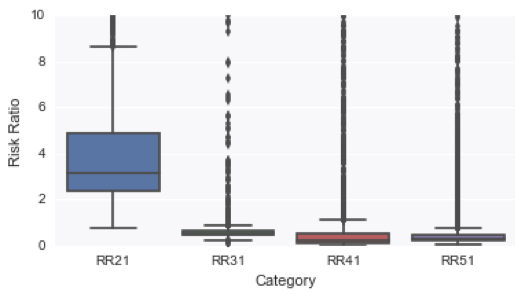}
  \caption{Posterior distribution of risk ratio of each category to category 1 (capped at 10).}
  \label{fig:fig2}
\end{figure}

The goals are to categorize users and the difference between different categories. We define the risk ratio (RR) of category $k (k\neq1)$ to category 1, which is defined as $RR_{k,1}=\frac{q_k}{q_1}$\. The risk ratio, also known as relative risk, is the ratio of the probability of event occurrence, which is often used to compare the probabilities of two group. As might be apparent from Figure \ref{fig:fig2}, $RR_{2,1}$ is estimated with propagation of the uncertainty. Seeing the MAP estimate and a 95\% credible interval of the distribution in Table \ref{tab:5}, category 2 users are about 2.6 times, at least 1.45 times with 95\% confidence, more likely to become active than category 1 users are.

\begin{table}[]
\caption{Posterior MAP, standard deviation, 95\% credible interval for the risk ratio toward category 1.}
  \label{tab:5}
\begin{tabular}{ccccc}
\hline
\textbf{}       & \textbf{MAP} & \multicolumn{1}{l}{\textbf{post. sd}} & \multicolumn{1}{l}{\textbf{Credible Interval}} & \multicolumn{1}{l}{\textbf{$\hat{R}$}} \\ \hline
\textbf{$RR_{21}$} & 2.6278                & 12.8555                               & (1.4539, 54.5580)                              & 1.0194                            \\
\textbf{$RR_{31}$} & 0.5001                & 5.6943                                & (0.3428, 5.7025)                               & 1.0219                            \\
\textbf{$RR_{41}$} & 0.0841                & 7.3732                                & (0.0059, 22.0108)                              & 1.0133                            \\
\textbf{$RR_{51}$} & 0.2563                & 5.9114                                & (0.1313, 14.8721)                              & 1.0157                           
\end{tabular}
\end{table}

\subsection{Estimated Survival Function and Hazard Function}
\begin{figure}[h]
  \centering
  \includegraphics[width=\linewidth]{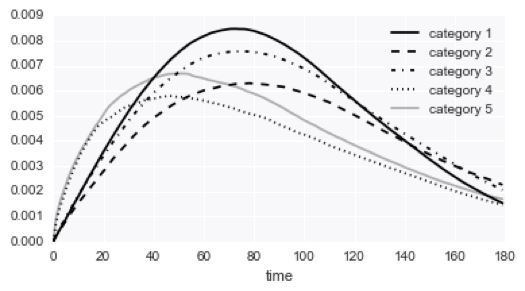}
  \caption{PDF of the Weibull distribution for each category with MAP estimates of shape and scale parameters.}
  \label{fig:fig3}
\end{figure}

To make the discussions in this subsection straightforward, all functions are drawn with MAP estimate.  Figure \ref{fig:fig3} presents the PDF of Weibull distribution for each category. The active users  follow this distribution. It is apparent that categories 4 and 5 have peaks at around time $t=45$. However, the distributions of categories 1, 2 and 3 peak around time $t=75$ and have fat tail. Therefore, the events of users in categories 4 and 5 more easily occur at first, but the events of categories 1, 2 and 3 are more likely to occur as time passes. This insight is consistent with our domain knowledge that users in categories 1, 2 and 3 are more prominent users for headhunters and that their popularity lasts longer than either Category 4 or 5.

\begin{figure}[h]
  \centering
  \includegraphics[width=\linewidth]{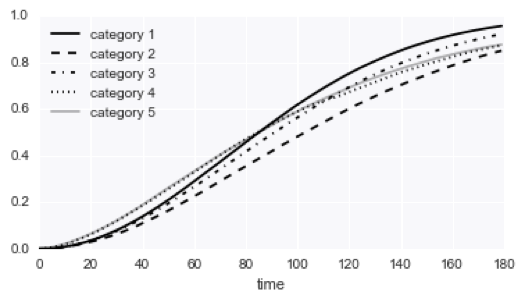}
  \caption{CDF of the Weibull distribution for each category with MAP estimates of shape and scale parameters.}
  \label{fig:fig4}
\end{figure}

\begin{figure}[h]
  \centering
  \includegraphics[width=\linewidth]{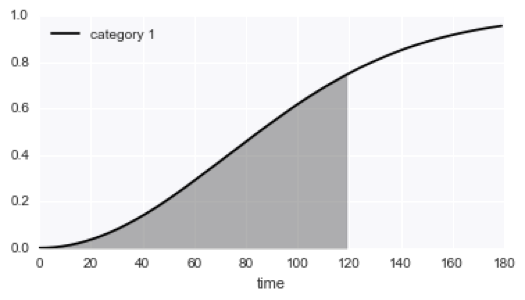}
  \caption{Estimated probability of event occurrence for category 1 until t=120 (shaded area).}
  \label{fig:fig5}
\end{figure}

Figure \ref{fig:fig4} shows CDF for each category. The total value of CDF in every category sums up to approximately 1.0 until the end because this Weibull distribution is applied to active users for whom events must happen in the future. Even if the user is an active user, some events will happen at time $t>180$. For example, the value of category 2 at $t=180$ is about 0.83. This can be inferred as indicating that the residuals, 17\% of the subjects' events, in category 2 will happen at time $t>180$. This inference derives from an assumption that the duration follows a parametric Weibull model of each category. Figure \ref{fig:fig5} also shows CDF for category 1. When taking the integral of CDF until time $t=120$, it is possible to estimate the probability of conversion until time $t=120$ (shaded area). One can calculate the whole conversion rate until time $t=120$ as shown below.
\begin{equation} \label{eq:17}
    q_k\times Pr_k(T\leq120)=q_k\times F_k(120)
\end{equation}

\begin{figure}[h]
  \centering
  \includegraphics[width=\linewidth]{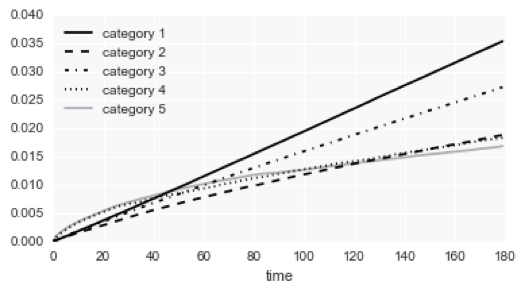}
  \caption{Hazard functions for each category with MAP estimates of shape and scale parameters.}
  \label{fig:fig6}
\end{figure}

Figure \ref{fig:fig6} presents hazard functions of the respective categories. In general, a hazard function implies the instantaneous rate of occurrence of the event. In categories 1, 2 and 3, the hazard function is increasing almost linearly, which means that events of these categories are more likely to occur as time goes by. Interestingly enough, the functions of categories 4 and 5 are convex upward, becoming horizontal with time.  Therefore, the rate of event occurrence remains stable in these categories during the time when the line is horizontal. 

\begin{figure}[h]
  \centering
  \includegraphics[width=\linewidth]{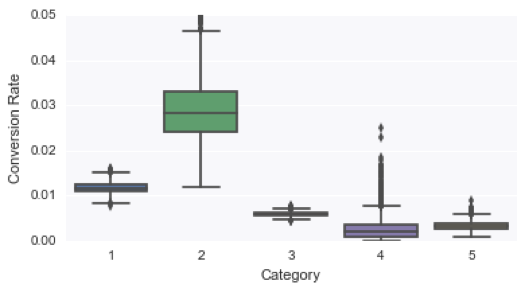}
  \caption{Posterior distribution of conversion rate until time$ t=120$ for each category.}
  \label{fig:fig7}
\end{figure}

Finally, we estimate the whole conversion rate until time t in Bayesian inference. One remarkable aspect of the proposed model is its interpretability. One can separate the event occurrence into Bernoulli trial and survival model, and assign an interpretation to each as presented above. Although the conversion rate was calculated with a MAP estimate in the examples presented above, one can also estimate it using Monte Carlo inference with all information of the posterior. Given each draw $\Theta^s=\{q^s,\lambda^s,\theta^s\}$ of MCMC sampling, we can sample any predictive quantity $g(\Theta^s)$. We define the generated quantity $g_k=q_k \times F_k(120)$ as the conversion rate, which describes the probability of event occurrence from user registration until time $t=120$. Figure \ref{fig:fig7} shows the posterior distributions of the quantity. One of the key findings of this research is shown in Figure \ref{fig:fig7}. It shows the posterior distributions of the quantity. Intuitively, we assumed that the category 1 is the most promising segment for our platform in terms of their educational background and age. However, our model suggests that category 2 is the most promising in terms of the expected conversion rate. This result helps us pivot around the target segments for our platform.

Table \ref{tab:6} presents a summary of the samples. As might be readily apparent, category 2 has the highest conversion rate with MAP estimate 0.0277. Moreover, it has a flat distribution. 


\begin{table}[]
\caption{Posterior MAP, standard deviation, 95\% credible interval of conversion rates for each category.}
  \label{tab:6}
\begin{tabular}{ccccc}
\hline
\textbf{}   & \textbf{MAP} & \textbf{post. sd} & \textbf{Credible Interval} & \textbf{$\hat{R}$} \\ \hline
\textbf{$g_1$} & 0.0117       & 0.0012            & (0.0094, 0.0141)           & 1.0029        \\
\textbf{$g_2$} & 0.0277       & 0.0065            & (0.0177, 0.0430)           & 1.0132        \\
\textbf{$g_3$} & 0.0058       & 0.0005            & (0.0050, 0.0068)           & 1.0362        \\
\textbf{$g_4$} & 0.0010       & 0.0025            & (0.0001, 0.0089)           & 1.0028        \\
\textbf{$g_5$} & 0.0030       & 0.0010            & (0.0016, 0.0055)           & 1.0001       
\end{tabular}
\end{table}

\section{CONCLUSION AND FUTURE WORK}
This paper extends classical survival analysis to a model that can accommodate extremely rare events occurrence in a natural way. We proposed a Bayesian Hierarchical Bernoulli-Weibull distribution Mixture model to avoid harmful effects of rare events affecting estimation.  To verify the validity of our model, experiments using real-world data were conducted. Our evaluation experiments based on WAIC and WBIC demonstrate that the proposed model outperforms the classical survival model and that the hierarchical structure further improves the model. Discussions of the estimation, uncertainty, and interpretation of the models were also given, which explains the characteristics and the difference between categories. One of the achievements of this research is revealing the gap between our intuition and the reality about our target segments. Our model tells us the target segments we should really focus on.  

For this study, we assume a Weibull distribution for the duration. Future work shall be undertaken to compare the Weibull model with different models such as exponential and log-logistic models.  Furthermore, future work will augment our study to apply this model to any other application domains.

\bibliography{KDD2019_survival}

\end{document}